\documentclass[number,preprint,review,12pt]{article}

\usepackage{amssymb}
\usepackage{amsmath}
\usepackage{comment}
\usepackage{siunitx}
\usepackage{multirow}
\usepackage{bm}
\usepackage{setspace}
\usepackage{authblk}
\usepackage[backend=biber,style=numeric,sorting=none]{biblatex}
\usepackage{setspace}
\usepackage{graphicx}

\title{A Closed-Form Solution for Kernel Adaptive Filtering} 

\author[a]{Benjamin Lyons Colburn} 
\author[b]{Luis G. Sanchez Giraldo}
\author[a]{Kan Li}
\author[a]{Jose C. Principe}

\affil[a]{\small Department of Electrical and Computer Engineering, University of Florida}
\affil[b]{ \small Department of Electrical and Computer Engineering, University of Kentucky}
\date{}
\begin{document}
\maketitle

\begin{abstract}
Unlike the conventional kernel adaptive filtering (KAF) approach of using a fixed kernel to define the Reproducing Kernel Hilbert Space (RKHS), this paper embeds the statistics of the input data in the kernel definition, obtaining a closed-form solution for nonlinear adaptive filtering. We call this solution the Functional Wiener Filter (FWF), and it is formally an extension of Parzen’s work on the autocorrelation RKHS to nonlinear functional spaces. We present a method for approximating the FWF in an explicit, finite-dimensional RKHS to model time series directly from realizations, which is less computationally demanding at test time than other KAF methods. We show that FWF outperforms KAF on a synthetic dataset that meets the conditions of the theory, and is comparable to other KAF algorithms for both a chaotic and real-world time series. We demonstrate how the difference equation learned by the FWF can be extracted, leading to possible applications in system identification.
\end{abstract}

\doublespacing
\section{Introduction}
\label{Intro}


While linear adaptive filters \cite{haykin2014adaptive} are well-established for applications including system identification, radar and sonar, active noise cancellation, channel equalization, and others, the inherent nonlinearity of most real-world systems motivates the development of nonlinear variants of these algorithms.
Despite their flexibility for modeling, in comparison with linear adaptive filtering algorithms, the theory for developing tractable and computationally efficient algorithms for nonlinear adaptive filtering is far from mature. In this work we bridge this gap, by extending the theory for minimum mean square error (MMSE) estimation developed by Emmanuel Parzen, to derive a closed-form solution for nonlinear adaptive filtering. 

Kernel adaptive filtering (KAF) \cite{Liu2010} offers a solution for nonlinear MMSE filtering. In general, KAF utilizes gradient search methods to find the optimal functional in the feature space of an RKHS, resulting in a nonlinear filter. Some examples of this approach are Kernel Least Mean Squares (KLMS) \cite{Liu2008} and Kernel Recursive Least Squares \cite{Engel2004}, which extend the Least Mean Squares (LMS) and Recursive Least Squares (RLS) algorithms, respectively. KAF is an active area of research, for a comparative study of some KAF methods, see \cite{Vaerenbergh2013}. More recent work on robust KAF includes \cite{Wang2022,He2023}. Methods for computationally efficient KAF can be found in \cite{Li2019, Wang2025}.

While effective and computationally feasible closed-form solutions for KAF have largely been absent, a few prior attempts have been made. For instance, in \cite{Pokharel2006} a nonlinear extension for the Wiener filter based on correntropy \cite{Santamaria2006} was discussed but had non-competitive performance and large test-time computational costs.

Other alternative methods for nonlinear MMSE filtering can be obtained in a Bayesian setting, in which the MMSE estimator is given by the posterior mean. 
A well-known Bayesian method for nonlinear MMSE estimation is Gaussian Process Regression (GPR) \cite{Rasmussen2006}. This method uses a Gaussian process (a random process where all joint distributions are Gaussian) to define a prior distribution over possible functions. Then the data is used to create maximum a posteriori estimators. Other related methodologies are based in the theory of splines \cite{Wahba1990}, where the problem of finding an optimal MMSE spline to fit data is presented in an RKHS. 

There are also deep learning based methods, such as LSTMs \cite{Hochreiter1997,Sherstinsky2020}, Temporal Convolutional Neural Networks \cite{Lea2016}, and Transformers \cite{Wen2023}. A recent survey of these methods is given in \cite{Miller2024}. However, the computational complexity of these models is not comparable with KAF, and the advances, theoretical and otherwise, made in this paper are most closely related to KAF and GPR. Therefore, we focus on comparison with KAF and GPR. 

The theory of optimal linear filtering based on MMSE estimation was initially introduced in the seminal works of Norbert Wiener \cite{Wiener1949} and Andrey Kolmogorov \cite{Shiryayev1992Interpolation}, but a closed-form solution for nonlinear filters has been elusive. In \cite{Parzen1961}, Parzen realized that because the autocorrelation function of the input random process is positive definite it can be used to define a data-dependent reproducing kernel Hilbert space (RKHS), where the optimal filter, in the MMSE sense, corresponds to a linear functional in the space. Parzen argued that embedding the statistical information of the autocovariance function into the inner product of the RKHS creates a natural space for statistical inference on random processes because conditional expectations with respect to a stationary input random process can be expressed as inner products in the RKHS. This claim of a ``natural" space is supported by the subsequent use of the autocovariance RKHS, denoted here as $\mathcal{H}_R$, to clarify and simplify many problems in statistical signal processing (see \cite{Kailath1971,Kailath1975,Duttweiler1973} for examples). 


In \cite{Mollenhauer2022}, the kernel autocovariance operators of stationary processes are theoretically studied and classical limit theorems as well as non-asymptotic error bounds under some ergodic assumptions are discussed. The autocovariance operators discussed in \cite{Mollenhauer2022} are closely related to the $U$ operator discussed in later sections of this paper. However, \cite{Mollenhauer2022} is focused on the analysis of the kernel autocovariance operator itself, while our work was independently developed and focuses on the extension of Parzen's work on MMSE filters. 

Our method is closely related to KAF in the way we construct a space of nonlinear functions of the input space, but it connects this space with Parzen's autocorrelation RKHS to obtain a closed-form solution. Although this closed-form solution requires an assumption that the data are stationary, it remains desirable for several important reasons. First, it gives insight into the system's performance allowing for more explicit characterization of design variable's effects on performance metrics. Second, it connects theory to practice, facilitating clearer principles for system design. Third, the optimal solution comes with guarantees; for example, we know that given our simplifying assumptions and our data, the solution is optimal, which aids in the analysis of the underlying system that creates the data. Finally, it may aid the development of optimal control laws. Therefore, the pursuit of a closed-form solution for nonlinear MMSE filters is worthwhile. 

Beyond being a closed-form solution, the FWF differs from other KAF methods in several consequential ways. First, while other KAF methods define a functional based solely on the amplitude of a random process, the FWF yields functionals over both time and amplitude values. Second, the FWF focuses on first building a data-dependent RKHS where the MMSE estimator can be given immediately. This fundamentally different approach to solving for the MMSE estimator makes the FWF distinct from other KAF methods. 

In summary, the main contributions of this work are the following. First, we introduce the theory to extend Parzen's MMSE solution in $\mathcal{H}_R$ to an RKHS that includes nonlinear functions of a random process, along with a method for computing this solution from realizations, yielding a closed-form, computationally efficient, and effective nonlinear MMSE estimator. Second, we provide a method for the practical implementation and use of a data-dependent nonlinear RKHS for signal processing applications. Third, we demonstrate experimentally that when the assumptions made in the closed-form solution are met, we outperform other kernel-based nonlinear filtering methods. Finally, we show that the optimal solution parameters can be interpreted as a nonlinear difference equation learned directly from the data. This extends possible applications of the method to system identification tasks and physics-based modeling in an RKHS. 

The remainder of the paper is organized as follows. First, a review of Parzen's linear MMSE solution in $\mathcal{H}_R$ is given. Next, we introduce the theory that extends this solution to include nonlinear functions. Then, we provide a practical method for computing this solution from realizations of input and target random processes. In sections \ref{sec:error}-\ref{sec:diffeqsection}, we give an analysis of the error given by the FWF, and show that the solution given by the FWF can be interpreted as a nonlinear difference equation. Finally, we compare the FWF with other KAF and nonlinear regression methods on two simulated time series and one real-world time series, and give concluding remarks. 
\section{MMSE Solutions in Data-Dependent RKHSs}
\subsection{Linear MMSE Solution}\label{subsec:linearMMSE}
Let $(\Omega,\mathcal{A},P)$ be a probability space with sample space $\Omega$, $\sigma$-algebra $\mathcal{A}$, and probability measure $P$.  A random process, $X = \{X_t, t \in T\}$, is a collection of random variables (r.v.) defined on, $(\Omega,\mathcal{A},P)$, along with an index set $T$ that is a compact subset of a separable metric space usually representing time. All random processes will be assumed to contain real-valued r.v.s. We denote a single r.v. within a random process with a capital letter subscripted with its index, $X_t$. Realizations from these random variables will be denoted as $x_t$.

In our treatment of the linear MMSE solution, we assume wide-sense stationarity of the processes involved. A definition of wide-sense stationarity can be found in \cite{Shiryayev1992sequences}.
Stationarity is a necessary assumption to practically estimate the quantities needed to solve the Wiener-Hopf equations from realizations of a random process \cite{Wiener1949}. A strictly stationary random process is a random process where the joint distributions (not just the second-order moments) do not change with shifts in time, which is required in our nonlinear MMSE solution.

We now review the linear MMSE solution given in \cite{Parzen1961} (also see \cite{Parzen1959}). Let the space of square-integrable r.v.s. defined on $(\Omega,\mathcal{B},P)$ be denoted as $L^2(\Omega,\mathcal{A},P)$. This is the space of all r.v.s., $W$ such that $ \Vert W \Vert_2 = \int_{\Omega} \vert W \vert^2 dP < \infty$.
The linear span of a random process, $X$, in $L^2(\Omega,\mathcal{A},P)$ is the smallest subspace of $L^2(\Omega,\mathcal{A},P)$ containing $X$ \cite{Parzen1959}. We can define this set by first defining the linear manifold $L(X_t,t\in T)$ as the set of all r.v.s. with the form $W = \sum_{i=1}^{n} a_i X_{t_i}$ with $a_i \in \mathbb{R}$, $t_i \in T$, and $n \in \mathbb{N}$. While $L(X_t,t\in T)$ is a linear manifold, it is not complete. We can complete this space by including the limits of all Cauchy sequences of elements in $L(X_t, t\in T)$. This complete set is the linear span of a random process in $L^2(\Omega,\mathcal{A},P)$, denoted as $L^2(X)$. Consider two r.v.s. $W,V \in L^2(X)$ where $W = \sum_{t \in T} a_t X_t$, $V=\sum_{s \in T}b_sX_s$, and $a_t,b_s \in \mathbb{R}$. The inner product in $L^2(X)$ can be written as
\begin{equation}
    \langle W,V \rangle_{L^2} = \mathbb{E}[WV] = \mathbb{E}\left[\sum_{s,t \in T}a_tb_sX_tX_s\right] = \sum_{s,t \in T}a_tb_s \mathbb{E}[X_tX_s].
    \label{innerL2X}
\end{equation}

Then with the positive semi-definite covariance function $R(s,t) = \mathbb{E}[X_tX_s]$, we see that the inner product between any two r.v.s. in $L^2(X)$ can be written in the RKHS whose kernel is defined by the covariance function of the random process $X$ as,
\begin{equation}
    \langle W, V \rangle_{L^2} =\sum_{s,t \in T} a_t b_s R(s,t)  =  \langle W^{'}, V^{'} \rangle_{\mathcal{H}_R}.
    \label{HRcongruence}
\end{equation}
Equation \eqref{HRcongruence} implies that $L^2(X)$ and $\mathcal{H}_R$ are congruent. This means that there exist a congruence mapping, an isomorphism $\psi(\cdot): L^2(X) \rightarrow \mathcal{H}_R$, such that $\langle W, V \rangle_{L^2(X)} = \langle\psi(W),\psi(V)\rangle_{\mathcal{H}_R}$. This congruence combined with Riesz Representation Theorem \cite{Riesz1907} guarantees that any linear functional over $L^2(X)$ has an exact representation in $\mathcal{H}_R$. Therefore, we can define equivalent solutions in either space. Since $L^2(X)$ contains all possible linear mapping functions over $X$, any linear MMSE solution has an exact representation in $\mathcal{H}_R$.

Suppose we are given the random process $X$ as input and $Z$ as the desired random process. As a consequence of Hilbert projection theorem, the linear MMSE solution can be given as the projection of $Z$ into $H_R$. In \cite{Parzen1961}, it is shown that the cross-covariance function is the linear MMSE solution in $\mathcal{H}_R$.
\noindent Therefore, the MMSE solution is the inner product between the cross covariance function $\rho$ with $X$ in $\mathcal{H}_R$ that is,
\begin{equation}
Z^* = \langle \rho,X\rangle_{\mathcal{H}_R} =\sum_{s,t \in T} X_t R(s,t)^{\dagger}\rho(s), \quad \rho(s) =\mathbb{E}[ZX_s],
\label{linearMMSE}
\end{equation}
\noindent where $\dagger$ is the Moore-Penrose pseudo inverse. In discrete time, this solution is equivalent to the Wiener solution, where the optimal impulse response is given by $w^*=\mathbf{R}^{\dagger}\bm{\rho}$, where $\mathbf{R}$ is the auto-covariance matrix and $\bm{\rho}$ is the cross-covariance vector. This demonstration shows that in the data-dependent RKHS, $\mathcal{H}_R$, there is no
 need to search for the linear MMSE solution. The solution is the cross-covariance function. The problem is that this solution is still linear in the input space. We now give the generalization of Parzen's work in \cite{Parzen1961} to include nonlinear functions.
\subsection{Nonlinear MMSE Solution}
\label{sec:nonlinearMMSE}
As before, let $X$ be a real-valued random process where each $X_t$ is a r.v. defined on a probability space $(\Omega, \mathcal{A}, P)$, and $T$ is a compact subset of a separable metric space. Let $\mathcal{F}_{\mathcal{X}} = \{f_x, x\in \mathcal{X}\}$ be a family of functions $f_x : \mathbb{R} \mapsto \mathbb{R}$ indexed by the elements of a compact set $\mathcal{X}$ such that $\mathbb{E}[\vert f_x(X_t) \vert^2] < \infty$ for all $x\in \mathcal{X}$ and $t \in T$. Note that if we let $x,y \in \mathcal{X} \subseteq \mathbb{R}$ and $f_x = \kappa(\cdot, x)$ where $\kappa(\cdot, \cdot)$ is a reproducing kernel, then these constraints will be met.
From \cite{Parzen1959}, we have $L_2(\Omega, \mathcal{A}, P)$ as the Hilbert space of r.v.s. in $(\Omega, \mathcal{A}, P)$ with finite second-order moments. We can define the set $f(X) = \{f_x(X_t), (x,t) \in \mathcal{X} \times T\}$ as a family of finite second-order random functions indexed by $x \in \mathcal{X}$ and $t \in T$. This set corresponds to the set of all r.v.s. that can be written in the form, $W = \sum\limits_{i=1}^{n_W}\sum\limits_{j=1}^{m_W} a_{ij} f_{x_i}(X_{t_j})$
for some positive integers $n_W$ and $m_W$. From the conditions defined above, we have that $W \in L^2(\Omega, \mathcal{A}, P)$.
By defining the inner product using the expected value as, 
\begin{equation}
\begin{split}
    \langle W, V \rangle_{L^2} = \mathbb{E}[WV] &= \sum\limits_{i=1}^{n_W}\sum\limits_{j=1}^{m_W}\sum\limits_{k=1}^{n_V}\sum\limits_{\ell=1}^{m_V} a_{ij}b_{k\ell} \mathbb{E}[f_{x_i}(X_{t_j}) f_{x_k}(X_{t_\ell})]\\ &= \sum\limits_{i=1}^{n_W}\sum\limits_{j=1}^{m_W}\sum\limits_{k=1}^{n_V}\sum\limits_{\ell=1}^{m_V} a_{ij}b_{k\ell} U(t_j, t_\ell, x_i, x_k),
\end{split}   
\label{nonlinearL2inner}
\end{equation}
\noindent we can form a linear manifold, $L(f_x(X_t), (x,t) \in \mathcal{X} \times T)$. Then, by adding the limits to all Cauchy sequences, we define the Hilbert space $L^2(f_x(X_t), (x,t) \in \mathcal{X} \times T)$, abbreviated as $L^2(f(X))$.
From \eqref{nonlinearL2inner} it is easy to see that for $s,t \in T$ and $x,y \in \mathcal{X}$ we can write the function $U(t,s,x,y)$ as $U((x,t),(y,s))$. This operator, $U: (\mathcal{X}\times T) \times (\mathcal{X}\times T) \mapsto \mathbb{R}$, is a positive semi-definite function, and is therefore the kernel for some RKHS, $\mathcal{H}_U$.

Equation (\ref{nonlinearL2inner}) suggests that $L^2(f(X))$ and $\mathcal{H}_U$ are congruent. Therefore, any MMSE solution in $L^2(f(X))$ has an exact representation in $\mathcal{H}_U$, and Parzen's solution in $\mathcal{H}_U$ is possible. The MMSE solution in $\mathcal{H}_U$ is the orthogonal projection of some desired r.v. $Z$ into the space. This projection can be written as $\rho(y,s) = \mathbb{E}[Zf_x(X_s)]$.
Finally, the MMSE solution in $\mathcal{H}_U$, which we call the Functional Wiener Filter (FWF), is given as
\begin{equation}
    Z^* = \langle \rho,X\rangle_{\mathcal{H}_U} = \sum_{s\in T}\sum_{t\in T} \sum_{x\in \mathcal{X}}\sum_{y \in \mathcal{X}} f_{x}(X_t) U((x,t),(y,s))^{\dagger}\rho(y,s).
    \label{nonlinearSolution}
\end{equation}

\subsubsection{Generalizing $\mathcal{H}_U$}
The space of functions we just described corresponds to functions of $n$ variables that can be expressed as sums of functions on individual variables, that is,
\begin{equation}\label{eq:sums_of_functions_single_variable}
g(x_1, x_2, \dots, x_n) = g_1(x_1) + g_2(x_2) + \cdots + g_n(x_n),
\end{equation} 
where $g_i \in \operatorname{span}\left\{f_x, x \in \mathcal{X} \right\}$. This is restrictive if we wish to construct nonlinear functions of more than one sample.  Using tap-delay embedding of the random process, we can generalize \eqref{eq:sums_of_functions_single_variable} to functions of vectors in $\mathbb{R}^D$. For a random process $X$ and a given set of relative times $\left[\tau_1,\tau_2, \dots,
\tau_{D-1}\right]$, we can define the tap-delay version $\bm{X}^D = \{\bm{X}^{D}_t, t \in T\}$, where $\bm{X}_t^D = [X_t, X_{t-\tau_1}, X_{t-\tau_2}, \cdots, X_{t-\tau_{D-1}}]^{\top}$. For simplicity, we assume that $t-\tau_i \in T$ for all $t \in T$ and all $i = 1, 2, \dots, D-1$.

\begin{figure*}[h]
    \centering
    \includegraphics[width=0.9\linewidth]{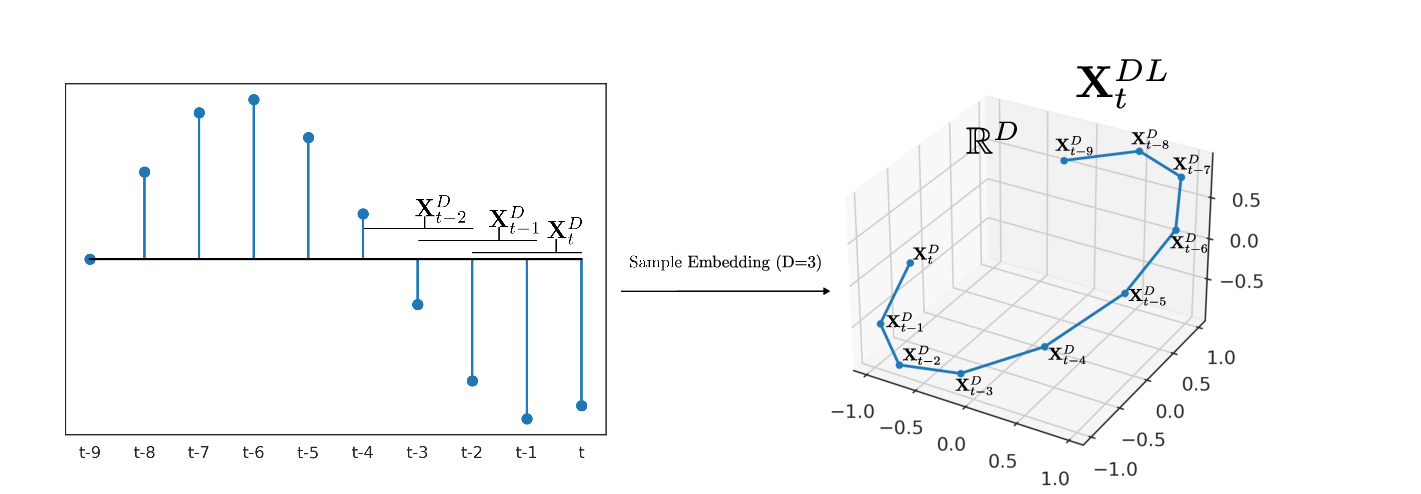}
    \caption{Visual depiction of sample embedding scheme.}
    \label{fig:sampleEmbed}
\end{figure*}

We now proceed just as before by defining the linear manifold $L(f_{x}(\bm{X}_{t}^D),(x,t) \in \mathcal{X} \times T)$ as the span of this family, where now $f_x : \mathbb{R}^D \mapsto \mathbb{R}$. Any r.v. in this manifold can be written as $ W = \sum_{i=1}^{n_W}\sum_{j=1}^{m_W} a_{ij}f_{x_i}(\bm{X}_{t_j} ^D)$.
Similarly, we denote the completion of this linear manifold by $L^2(f_x(\bm{X}_{t}^D),(x,t) \in \mathcal{X} \times T)$ abbreviated as $L^2(f(\bm{X}^D))$. 
Furthermore, we can extend the above notation to build nested sets. For example, if $T$ is the set of integers, we can choose a positive integer $L > 0$ to build the nested set $\bm{X}_t^{DL} = \{\bm{X}_t^D,\bm{X}_{t-1}^D...,\bm{X}_{t-(L-1)}^D\}$ and $\bm{X}^{DL} = \{\bm{X}_t^{DL}, t \in T\}$. 
We will refer to $D$ as the sample embedding size, and $L$ as the window size. By varying $D$ and $L$, the combinations of r.v.s. in $X$ over which the functions in $\mathcal{H}_U$ are defined can be adjusted. Figure \ref{fig:sampleEmbed} gives a visual depiction of the sample embedding scheme. 

In later sections, these nested sets will give the building blocks for creating a spectrum of spaces with different levels of generality and computation requirements. An overview of the trade-offs controlled by the hyperparameters of the FWF is given in section \ref{sec:hyperparameters}.



\section{Computing the MMSE Solution}

The sections above show theoretically how to extend Parzen's idea of a MMSE in a data-dependent and potentially universal RKHS. Now, we give one method for practically computing this nonlinear MMSE solution. We first define a congruence which shifts the domain of $U(s,t,x,y)$ where $x$ and $y$ are from a potentially uncountable set, to a countable domain. We then introduce an explicit finite-dimensional RKHS that approximates the Gaussian RKHS. Finally, we approximate the solution given in section \ref{sec:nonlinearMMSE} in this finite-dimensional RKHS.
\subsubsection{Mercer's theorem and a simple congruence}\label{sec:MercerDecomp}
For the case where we use a positive definite kernel $\kappa$ to define $f_x:=\kappa(\cdot, x)$ with $x\in \mathcal{X}$, where $\mathcal{X}$ is a compact space (for instance a closed interval in $\mathbb{R}^d$), we can define a congruence based on Mercer's theorem. This congruence can be used to change the domain of $U(s,t,x,y)$ where $x$ and $y$ are from a potentially uncountable set, to a countable domain. First, note that Mercer's theorem allows us to decompose the kernel as $\kappa(x,y) = \sum\limits_{m=1}^{N_{\mathcal{H}_{\kappa}}}\lambda_m\psi_m(x)\psi_m(y)$, where $N_{\mathcal{H}_{\kappa}}$ is either finite or countably infinite. Then $L^2(\kappa(X))$ is congruent with $L^2(\psi_m(X_t), (t,m) \in T \times \mathbb{N})$ and consequently also congruent with $\mathcal{H}_{\bm{U}}$ where $\bm{U}:(T\times \mathbb{N}) \times (T\times \mathbb{N})\mapsto \mathbb{R}$ is a positive definite kernel defined as,
\begin{equation}
    \bm{U}((t,m), (s,n)) = \mathbb{E}\left[ \psi_m(X_t)\psi_n(X_s)\right],
\end{equation}
\noindent where $\bm{U}(s,t,n,m)$ is now defined only over countable sets.

\subsection{The Explicit Finite-Dimensional Approximation for the Gaussian RKHS}
In \cite{Li2019}, an explicit mapping function based on the Taylor expansion of the Gaussian kernel ($G(\cdot,\cdot)$) is used to give a finite-dimensional approximation of the feature space specified by the Gaussian kernel. A full derivation of this explicit mapping function can be found in \cite{cotter2011}. This is just one way of creating an explicit feature space; another notable technique employs Random Fourier features (RFF) \cite{Rahimi2007}, but it has higher computational costs than the Taylor expansion-based method. We will refer to this explicit finite-dimensional Hilbert Space as $\mathcal{H}_{S}$, with kernel $S(\cdot,\cdot): \mathbb{R}^D \times \mathbb{R}^D \rightarrow \mathbb{R}$. The explicit mapping function given in \cite{cotter2011} (also see \cite{Yang2004nips}) is written as
\begin{equation}
   \phi_{k,j}(x) = e^{-\frac{\Vert \mathbf{x} \Vert^2}{2 \sigma^2}} \frac{1}{\sigma^k\sqrt{k!}} \prod_{i=1}^{k}x_{j_i},
   \label{explicitMapping}
\end{equation}
where $x \in \mathbb{R}^D$, and $j \in [D]^k$ enumerates over all selections of $k$ coordinates of $x$ (allowing repetitions and enumerating over different orderings of the same coordinates). For instance, the set $[2]^3$ consists of eight $3$-tuples, namely,  $\left(1, 1, 1\right)$, $\left(1, 1, 2\right)$, $\left(1, 2, 1\right)$, $\left(1, 2, 2\right)$, $\left(2, 1, 1\right)$, $\left(2, 1, 2\right)$, $\left(2, 2, 1\right)$, and $\left(2, 2, 2\right)$.\\
The finite rank approximation of the Gaussian kernel is then obtained by truncating the Taylor series expansion to the first $K$ terms, 
\begin{eqnarray}
\nonumber    S(x,x^{\prime}) & = & \left\langle\phi(x),\phi(x^{\prime})\right\rangle_{\mathcal{H}_S} =  \sum\limits_{k=0}^K\sum\limits_{j\in[D]^k}\phi_{k,j}(x)\phi_{k,j}(x^{\prime})\\ 
    & = & e^{-\frac{\Vert x \Vert^2}{2 \sigma^2}}e^{-\frac{\Vert x^{\prime} \Vert^2}{2\sigma^2}}\sum\limits_{k=0}^K\frac{\left(x^{\top}x^{\prime}\right)^k}{\sigma^{2k}k!}= \sum_{m=1}^{M} \phi_m(x)\phi_m(x^{\prime}).
\label{Sxx}
\end{eqnarray}
The last expression in (\ref{Sxx}) is given by collecting like monomials and flattening $\phi_{k,j}(x)$ into a vector of size $M=\binom{D+K}{K}$, where D is the dimension of the input vectors and K is the truncation point. We will use this simplified representation of $\phi(x)=\{\phi_m(x)\}_{m=1}^{M}$ from now on.

\par Since the Gaussian kernel over a closed bounded interval is a Mercer kernel, the representations based on the full Taylor expansion and the eigendecomposition of the integral operator induced by the kernel (Mercer's theorem) are equivalent. Therefore, the congruent relationship detailed in \ref{sec:MercerDecomp} applies to the Taylor expansion, and the closure of the span of $\{\phi_{m}(X_t), (m,t) \in \mathbb{N}\times T\}$ contains $L^2(\{G_x(X_t), (x,t) \in \mathcal{X}\times T\})$. Truncating this series yields a family of functions, $\{\phi_m(X_t), (m,t) \in [1,M] \times T\}$, where we can approximate any $f_x = G(\cdot, x)$ by a finite superposition, $f_x(x^{\prime}) \approx \sum\limits_{m=1}^M \phi_{m}(x)\phi_{m}(x^\prime)$.

In previous works \cite{Li2019,Rahimi2007,Yang2004nips} an explicit mapping function is used to decouple model size from the number of training samples. While our method inherits this as a strength of using an explicit approximation, the main advantage of the explicit feature space in the context of this work is that it simplifies the design of linear operators because we can represent them as finite-dimensional matrices. This allows for a practical method for computing the closed-form solution detailed in \ref{sec:nonlinearMMSE}. The drawback of the finite-dimensional RKHS is that it is no longer universal. However, it is shown in \cite{Li2019}, and in later sections (also see supplementary materials), that effective finite-rank approximations can be obtained with just a few features in the expansion.

\subsection{Covariance kernel for the approximate MMSE Solution with the explicit feature map}\label{sec:MMSEComputation}
The solution in section \ref{sec:nonlinearMMSE} requires the definition of a family of r.v.s. with finite second-order moments. Our truncated approximation that computes an explicit feature map provides us with an alternative family of r.v.s. $\{\phi_{m}(\bm{X}^{D}_t), (m,t) \in [1,M] \times T\}$, abbreviated as $\phi(\bm{X}^D)$ as our family of functions with finite second-order moments. Now, we follow the steps detailed in section \ref{sec:nonlinearMMSE}. First, we define the covariance kernel, $\bm{U}(s,t,m,n)$ and the inner product in $\mathcal{H}_{\bm{U}}$. Then we demonstrate how to compute the cross-covariance function, yielding the nonlinear MMSE solution in $\mathcal{H}_{\mathbf{U}}$.   

With our family of functions $\phi(\bm{X}^D)$, the covariance kernel $\bm{U}$ is given as,
\begin{equation}
    \bm{U}(t,s,m,n) = \mathbb{E}[\phi_{m}(\bm{X}_t^D)\phi_{n}(\bm{X}_s^D)], \quad t,s \in T \quad m,n \in [1,M].   
\label{Ufunc}
\end{equation}
We can represent any random variable $W \in L^2(\phi(\bm{X}^D))$ as,
\begin{equation}\label{generalL2_truncated}   
W   = \sum_{q=1}^{m_W} \sum\limits_{m=1}^{M} \sum_{i=1}^{n_W} a_{iq} \phi_m(x_i)\phi_{m}(\bm{X}_{t_q}^D) = \sum_{q=1}^{m_W} \sum\limits_{m=1}^{M}A_{q, m}\phi_{m}(\bm{X}_{t_q}^D),
\end{equation}
\noindent where $A_{q,m} = \sum_{i=1}^{n_W}a_{iq}\phi_m(x_i)$. The inner product is given by,
\begin{equation}
\begin{split}
    \langle W, V \rangle_{L^2} = \mathbb{E}[WV] &= \sum_{q=1}^{m_W} \sum\limits_{m=1}^{M}\sum\limits_{p=1}^{m_V}\sum\limits_{n=1}^{M} A_{q,m}B_{p,n} \mathbb{E}[\phi_{m}(X_{t_q}) \phi_{n}(X_{t_p})]\\ &= \sum_{q=1}^{m_W} \sum\limits_{m=1}^{M}\sum\limits_{p=1}^{m_V}\sum\limits_{n=1}^{M} A_{q,m}B_{p,n} \bm{U}(t_q, t_p, m, n).
\end{split}   
\label{nonlinearL2inner_truncated}
\end{equation}
Assuming strict stationarity of $X$, we have that $\bm{U}(t,s,m,n) = \bm{U}(t-\tau,s-\tau,m,n)$. If we pick a set of $L$ relative times to $t \in T$, the joint statistics of the set of random vectors $[\bm{X}^{D}_t, \bm{X}^{D}_{t-1}, \cdots, \bm{X}^{D}_{t-(L-1)}]$ are the same as $[\bm{X}^{D}_s, \bm{X}^{D}_{s-1}, \cdots, \bm{X}^{D}_{s-(L-1)}]$. Then we can represent the relative time information along with the feature index as a matrix $\mathbf{U} \in \mathbb{R}^{(M\cdot L)\times(M \cdot L)}$.  
\noindent The details of exactly how to construct this matrix and its relation to $U(s,t,x,y)$ are given in \ref{A1}. 

Finally, we can define the projection of a desired r.v. $Z$ into $\mathcal{H}_{\bm{U}}$ using the cross-covariance function, $\tilde{\rho}(t,m) = \mathbb{E}[Z\phi_{m}(\bm{X}^D_t)]$.
The feature space representation of this function can be given as,
\begin{equation}
    \bm{\rho} = \mathbb{E}[Z\phi(\bm{X}_t^{DL})],  \quad \phi(\bm{X}^{DL}_t) :=
\begin{bmatrix}
\phi(\bm{X}^D_t)\\
\vdots\\
\phi(\bm{X}^D_{t-(L-1)})
\end{bmatrix}   \in \mathbb{R}^{M \cdot L}.
    \label{crosscorrvec}
\end{equation}
Then the MMSE in $\mathcal{H}_{\bm{U}}$ is 
\begin{equation}
    \hat{Z} = \langle\phi(\mathbf{X}_t),\tilde{\rho}\rangle_{\mathcal{H}_{\bm{U}}}= \phi(\bm{X}_t^{DL})^\top\mathbf{U}^{\dagger} \bm{\rho}.
    \label{nonlinearMMSEcalc}
\end{equation}
Notice that this equation has the same form as the linear MMSE solution given in equation (\ref{linearMMSE}) except that the number of dimensions is larger for the nonlinear case. While the number of dimensions in the linear MMSE solution is related only to time, the number of dimensions in the nonlinear solution is related to both time and dimensionality of $\phi(\cdot)$ as a result of the RKHS we employ. See Table \ref{tab:RKHSs} for a comparison between all RKHSs introduced thus far.
\begin{singlespace}
\begin{table}[h]
\resizebox{\columnwidth}{!}{
\small
    \begin{tabular}[width = \textwidth]{c|c|c|c|c}
       \textbf{RKHS}  & \textbf{Domain of Kernel Function} & \textbf{Nonlinear}&\textbf{Data-dependent}&\textbf{Includes Notion of Time}\\
      \hline
        $\mathcal{H}_R$ & $T \times T$ && \checkmark & \checkmark\\
        \hline
        $\mathcal{H}_\kappa$ & $\mathbb{R}^D \times \mathbb{R}^D $ & \checkmark & & \\
        \hline
        $\mathcal{H}_S$ & $\mathbb{R}^D \times \mathbb{R}^D $ & \checkmark & & \\
        \hline
        $\mathcal{H}_U$ & $(\mathbb{R}^D \times T) \times (\mathbb{R}^D \times T)$ & \checkmark & \checkmark & \checkmark\\
        \hline
        $\mathcal{H}_{\bm{U}}$ & $(\mathbb{N} \times T) \times (\mathbb{N} \times T)$ & \checkmark & \checkmark & \checkmark\\
        
    \end{tabular}
    }
    \caption{Brief description of the different RKHSs used so far. Note $\mathcal{H}_{\bm{U}}$ is in practice truncated.}
    \label{tab:RKHSs}
\end{table}
\end{singlespace}
\subsection{Error Analysis}\label{sec:error}

The FWF solution has several error sources. First, since calculations occur in a finite-dimensional RKHS, $\mathcal{H}_S$ and transitively $\mathcal{H}_U$ are only universal as K approaches infinity, D equals the true model order, and L equals one. Error bounds for $S(\cdot, \cdot)$'s approximation to the universal Gaussian kernel are given in \cite{cotter2011}. Second, the proper selection of the kernel size remains necessary, affecting precision with finite training datasets. Third, by quantifying the joint distribution between sample pairs projected into the RKHS and taking their mean, we implicitly assume strict stationarity. Any deviations from this assumption may introduce error. Finally, numerical error can arise from the conditioning of the $\mathbf{U}$ matrix.

It may seem that all these approximations are challenging to quantify; however, the optimal solution provides a direct means for evaluating the MSE. In fact, since the optimal solution is an orthogonal projection, we can calculate the cumulative approximation MSE by simply measuring the distance between the projection of $Z$ into $\mathcal{H}_{\bm{U}}$ (i.e. $\tilde{\rho}$) and $Z$ itself.
\begin{equation}
   \mathbb{E}[|Z - \mathbb{E}[Z|L^2(\phi(\bm{X}^{D}))]|^2]  = \mathbb{E}[|Z|^2] - \langle \tilde{\rho}, \tilde{\rho}\rangle_{\mathcal{H}_{\bm{U}}}. 
   \label{TheoErrorNonlinear}
\end{equation}
Since the projection of $Z$ in $\mathcal{H}_{\bm{U}}$ is orthogonal, the difference between the power of the desired response and the norm squared of its projection in $\mathcal{H}_{\bm{U}}$ is the expected MSE. Therefore, it is quite easy to select the hyperparameters of the model ($K$, $L$, $D$, and kernel size) to meet the minimum error specification. This is markedly different from KAF and other filtering methods, where we have to evaluate the MSE on the training set by directly comparing model outputs with desired responses for different hyperparameters. Discounting numerical error, all the sources of error mentioned above are combined in the inner product calculation of the space $\mathcal{H}_{\bm{U}}$. 

\begin{figure}[h]
    \centering
    \makebox[\textwidth]{\includegraphics[width=0.9\textwidth]{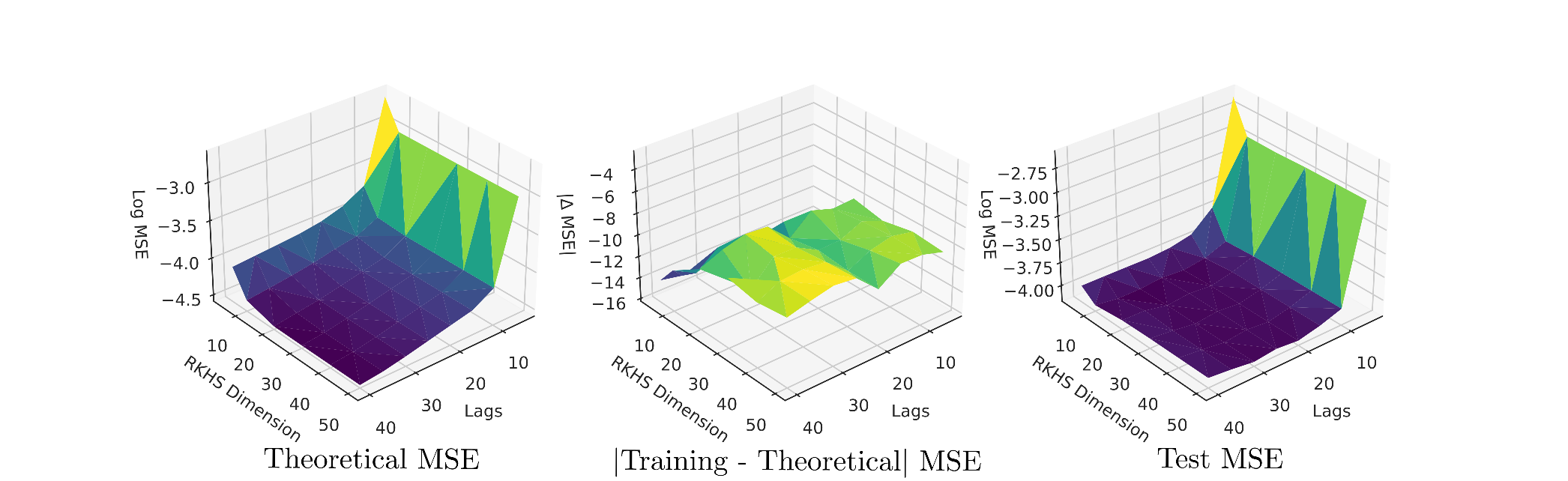}}
    \caption{ $|$Training - Theoretical$|$ MSE in log scale (middle), Theoretical MSE (left), and Test MSE (right) as a function of $M$ and $L$ for prediction of Mackey-Glass time series.}
    \label{fig:TrainTheo3D}
\end{figure}

Figure \ref{fig:TrainTheo3D} shows the theoretical MMSE, the absolute difference between theoretical and empirical MSE in the training set, and the test set MSE as a function of the FWF hyperparameters for prediction on the nonlinear chaotic Mackey-Glass time series introduced in \ref{sec:MackeyGlass}. These empirical results confirm that the theoretical MMSE closely matches empirical error across a wide range of hyperparameters, and test set MSE largely follows training set MSE. Since the FWF is a linear model in the RKHS, techniques for model order estimation \cite{Stoica2004} can be applied to improve generalization, but we leave this to future work.   

\subsection{The Functional Wiener Filter as a Difference Equation} \label{sec:diffeqsection}
Similar to the linear Wiener Filter, the Functional Wiener Filter can be interpreted as the optimal solution under the assumption of a difference equation with a specific form. 
The difference equation assumed by the Functional Wiener Filter is similar to a linear FIR filter except we replace scalar multiplication with nonlinear functions. Due to sample embeddings, these nonlinear functions may be defined over multiple samples rather than single samples. 
\begin{equation}
    \hat{z_t} = \sum^{L-1}_{\tau=0} f_{\tau}(\mathbf{x}_{t-\tau}^{D}) \quad f_\tau \in \mathcal{H}_S,\mathbf{x}_t^D \in \mathbb{R}^D
    \label{nonlineardiffeq}
\end{equation}
The easiest way to see how the FWF defines a difference equation is by first calculating $\mathbf{w}^* = \mathbf{U}^{\dagger}\bm{\rho}$. Then the FWF solution can be written as,
 \begin{equation}
     Z_t = \phi(\bm{X}_t^{DL})^\top \mathbf{U}^{\dagger}\bm{\rho} = \phi(\bm{X}^{DL}_t)^\top \mathbf{w}^*.
     \label{diffeqexpalined}
 \end{equation}
\noindent The expression $\phi(\bm{X}^{DL}_t)^\top \mathbf{w}^*$ in equation \ref{diffeqexpalined} can be written as,
\begin{equation}
   \hat{z}_t  = \sum_{\tau=0}^{L-1} \phi(\mathbf{x}^{D}_{t-\tau})^\top \mathbf{w}_{M\tau:(M\tau)+(M-1)}^* = \sum_{\tau=0}^{L-1}f^{*}_\tau(\mathbf{x}_{t-\tau}^{D}).
   \label{diffeqfinal}
\end{equation}
\noindent Therefore the subvectors, $\mathbf{w}_{M\tau:(M\tau)+ (M-1)}^*$, are the feature space representations of $\{f^{*}_{\tau}\}_{\tau=0}^{L-1}$.
Note that $\mathcal{H}_S$ becomes universal on $\mathbb{R}^D$ when $K$ goes to infinity. Every solution found using the Functional Wiener Filter follows this form. This highlights a distinction between the FWF and other KAF methods found in \cite{Liu2008,Liu2009,Engel2004,Liu2010}. The FWF is defined over $\mathbb{R}^D \times T$ (time and space) rather than just over $\mathbb{R}^D$. In the experimental section, we will demonstrate how this interpretation can be used to extract a difference equation from data. To our knowledge, this type of interpretation is not possible with any other KAF method. 

\subsection{Hyperparameters and Trade-offs}\label{sec:hyperparameters}
The hyperparameters of the FWF are sample embedding size ($D$), window size ($L$), truncation point ($K$), kernel size ($\sigma$), and regularization parameter (implicit in the Moore-Penrose pseudo inverse). The kernel size plays the same role as in other KAF methods. See \cite{Vega2019} for a discussion on kernel size selection. The regularization parameter also plays a standard role, controlling the conditioning for the inversion of $\mathbf{U}$.

From a high level, the roles of $D$, $L$, and $K$ are similar; they increase model capacity and generalize $\mathcal{H}_{\mathbf{U}}$. The specific roles of $D$ and $L$ are best understood using \eqref{diffeqfinal}. $D$ controls the domain of each function, $f^*_{\tau}$, while $L$ controls the memory depth of the system. $K$ affects model capacity by controlling the highest degree considered in our truncated Taylor series, which ultimately affects the hypothesis space for each $f_\tau^{*}$. If $D$, $L$, or $K$ are too small, then a satisfactory solution may not exist in $\mathcal{H}_{\bm{U}}$. Conversely, if they are too large, you unnecessarily increase the computational complexity, susceptibility to noise, and amount of data necessary for estimating the FWF well. Therefore, practitioners should aim to use the smallest values for $D$, $L$, and $K$ that permit the FWF to fit the data well.  

\section{Experiments and Simulations}
We now test the FWF solution in two important applications: nonlinear mapping of one time series into another (as required in system identification) and nonlinear time series prediction. The models used for comparison are the linear Wiener Filter (WF) \cite{Wiener1949}, Gaussian Process Regression (GPR) \cite{Williams1995}, Kernel Least-Mean Squares (KLMS) \cite{Liu2008}, Extended Kernel Recursive Least-Squares (KRLS) \cite{Liu2009}, Kernel Ridge Regression (KRR) \cite{Hoerl2000}, and Augmented Space Linear Model (ASLM) \cite{Qine2020}. The Gaussian kernel was used for all kernel methods.

For each experiment, a grid search was conducted across the hyperparameters of each model. The best results for each method are presented. For brevity, we give the finer details of the searches and final hyperparameter settings in the supplementary materials. An additional experiment on forecasting the Lorenz system is also included in the supplementary materials.
\begin{singlespace}
\begin{table}[h]
    \centering
    \small
    \begin{tabular}{ c|c|c}
 \textbf{Method} & \textbf{Training} & \textbf{Evaluation}\\
 \hline
 FWF & $\mathcal{O}(M^2 L^2 N) +\mathcal{O}(M^3L^3)$&$\mathcal{O}(ML)$ \\ 
 KLMS & $\mathcal{O}(N)$ &$\mathcal{O}(N)$ \\  
 KRLS &$\mathcal{O}(N^2)$ &$\mathcal{O}(N)$\\
 ASLM & $\mathcal{O}(L^2N)$ &$\mathcal{O}(L)+\mathcal{O}(log(N))$\\
 KRR & $\mathcal{O}(N^3)$& $\mathcal{O}(N)$\\
 GPR & $\mathcal{O}(N^3)$& $\mathcal{O}(N)$\\
 WF & $\mathcal{O}(L^2N)$ & $\mathcal{O}(L)$\\
\end{tabular}
    \caption{Computational Complexity Comparison for both training and evaluation: $M = \binom{D+K}{K}$ (number of dimensions in $\mathcal{H}_{S}$), N (number of training samples), L (window size)}
    \label{tab:ComplexTable}
\end{table}
\end{singlespace}
Table \ref{tab:ComplexTable} shows a comparison of the computational complexity between the different methods. The hyperparameters that affect the computational complexity of the FWF are $D$, $K$, and $L$.  The FWF training complexity is the highest, proportional to the cube of the product of window size and dimension; however, the FWF is a batch method, making the computation in the training step parallelizable. In practice, it is often the case that $N>>ML$, so the $\mathcal{O}(M^2L^2N)$ will dominate the training complexity, which is similar to the linear case. While training complexity can be large, test time complexity is untethered from N, similar to the WF, which is a great computational advantage for low-power computation.


\subsection{Demonstration of the Difference Equation Interpretation}
The interpretation given in section \ref{sec:diffeqsection} is now applied to identify the underlying functions that generate a time series. The simulated data is generated as follows. The input to the system ($x$) is white Gaussian noise with i.i.d. samples drawn from the distribution $\mathcal{N}(0,\pi)$. The system output, $z$, is obtained via the nonlinear mapping 
\begin{equation}
\small
    \begin{split}
        z_t  =& 0.5 \tanh(x_t)^2 + \sin(x_{t-1})^3 + 0.5 \tanh(x_{t-2})^3\\ 
        &+ 0.2 \sin(x_{t-3})^2 + 0.75 \tanh(x_{t-4})^2.
    \end{split}
    \label{SSsignal}
\end{equation}
\begin{figure*}[h]
    \centering
     \makebox[\textwidth]{\includegraphics[width=0.9\textwidth]{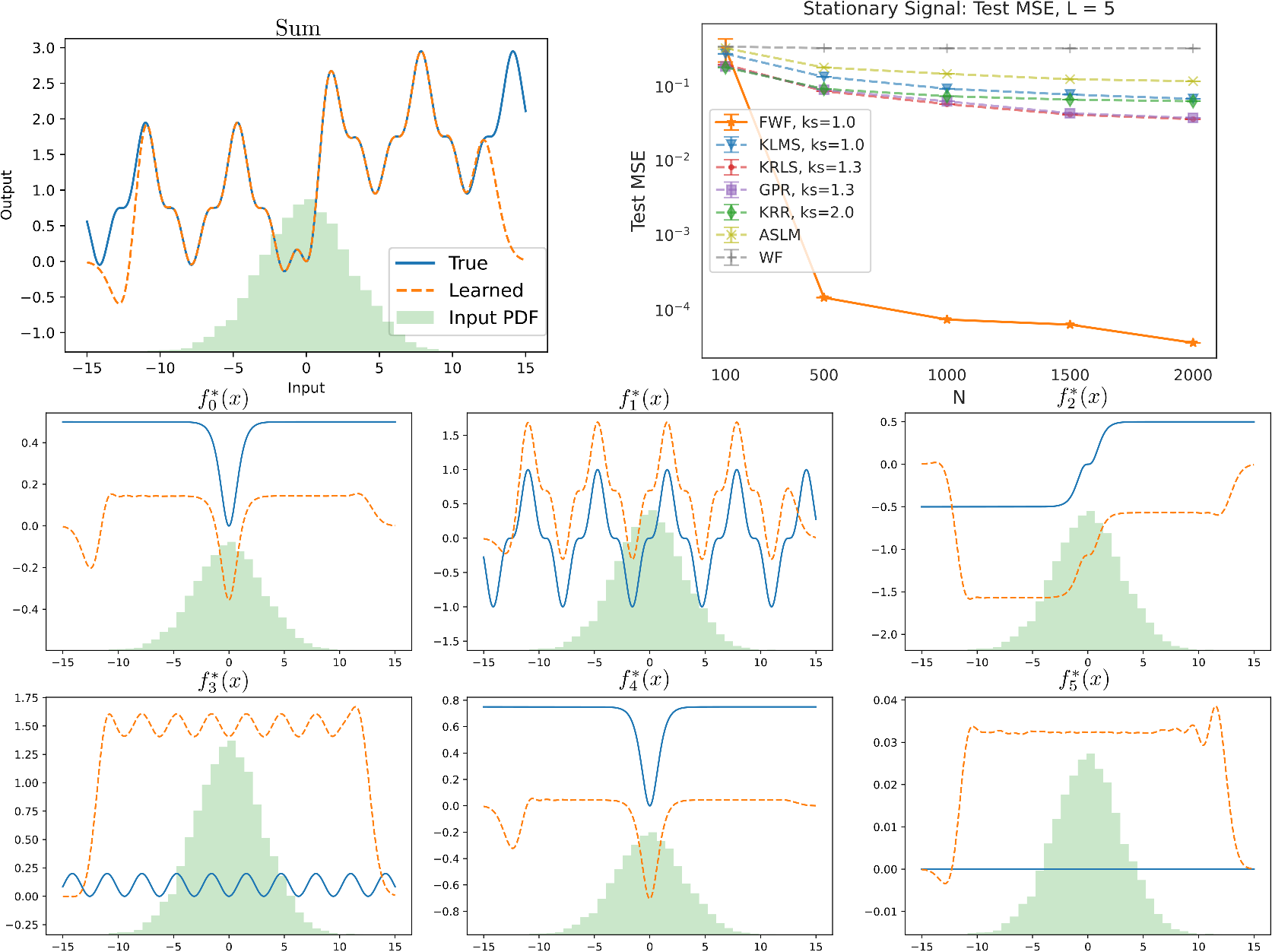}}
    \caption{Visualization of the underlying functions learned by the FWF on the strictly stationary task(bottom) and their sum(top left). Comparison of test set MSE as a function of the number of training samples(N). (top right). ks is the kernel size used for each method.}
    \label{fig:FWFVis}
\end{figure*}
\
\noindent The task is to estimate the mapping from $x$ to $z$. Note this is a standard setup for system identification. 

The bottom half of Figure \ref{fig:FWFVis} shows the set of functions $\{f^*_\tau(\cdot)\}^{L-1}_{\tau=0}$ found by the FWF. The green histograms show the p.d.f of the input signal used for training, scaled for visualization. We observe that in regions covered in the training set, the functions learned by the FWF are biased versions of the true functions given in equation (\ref{SSsignal}). However, when these biased versions are summed together, they converge to the true difference equation given in (\ref{SSsignal}). Centering the covariance in RKHS will compensate for the bias. It is noteworthy that the FWF outputs remain constant for values of $\tau$ exceeding the true memory depth of the system. 

The top right plot in Figure \ref{fig:FWFVis} compares the performance of the FWF with the other methods mentioned above. Each method was tested with five independent training and testing windows at each value of $N$. The average test set MSE is shown. The FWF provides a clear performance boost over the other methods on this task. This suggests that if the model assumptions hold, then the FWF's performance is better than the other nonlinear filtering methods. Moreover, we can plot the underlying functions learned by the FWF at each lag and can interpret the results as a difference equation.
\subsection{Mackey-Glass Prediction}\label{sec:MackeyGlass}
The Mackey-Glass time series is a chaotic nonlinear time series governed by the equation,
\begin{equation}
    \frac{dx_t}{dt} = \frac{a\theta^nx_{t-\tau}}{\theta^n+x_{t-\tau}^{n}} - bx_t.
    \label{MG}
\end{equation}
This time series was introduced in \cite{Mackey1977} as a model capable of producing nonlinear chaotic behavior similar to respiratory and hematopoietic diseases, and is commonly used to test time series prediction methods (see \cite{Wang2022,Liu2010}). We use the values $a=0.2$, $b=0.1$, $n=10$, $\theta=1$ and $\tau=30$. The time series is then discretized with a sampling period of 6 seconds using the fourth-order Runge-Kutta method.
\begin{figure*}[h]
    \centering
    \makebox[\textwidth][c]{\includegraphics[width=0.85\textwidth]{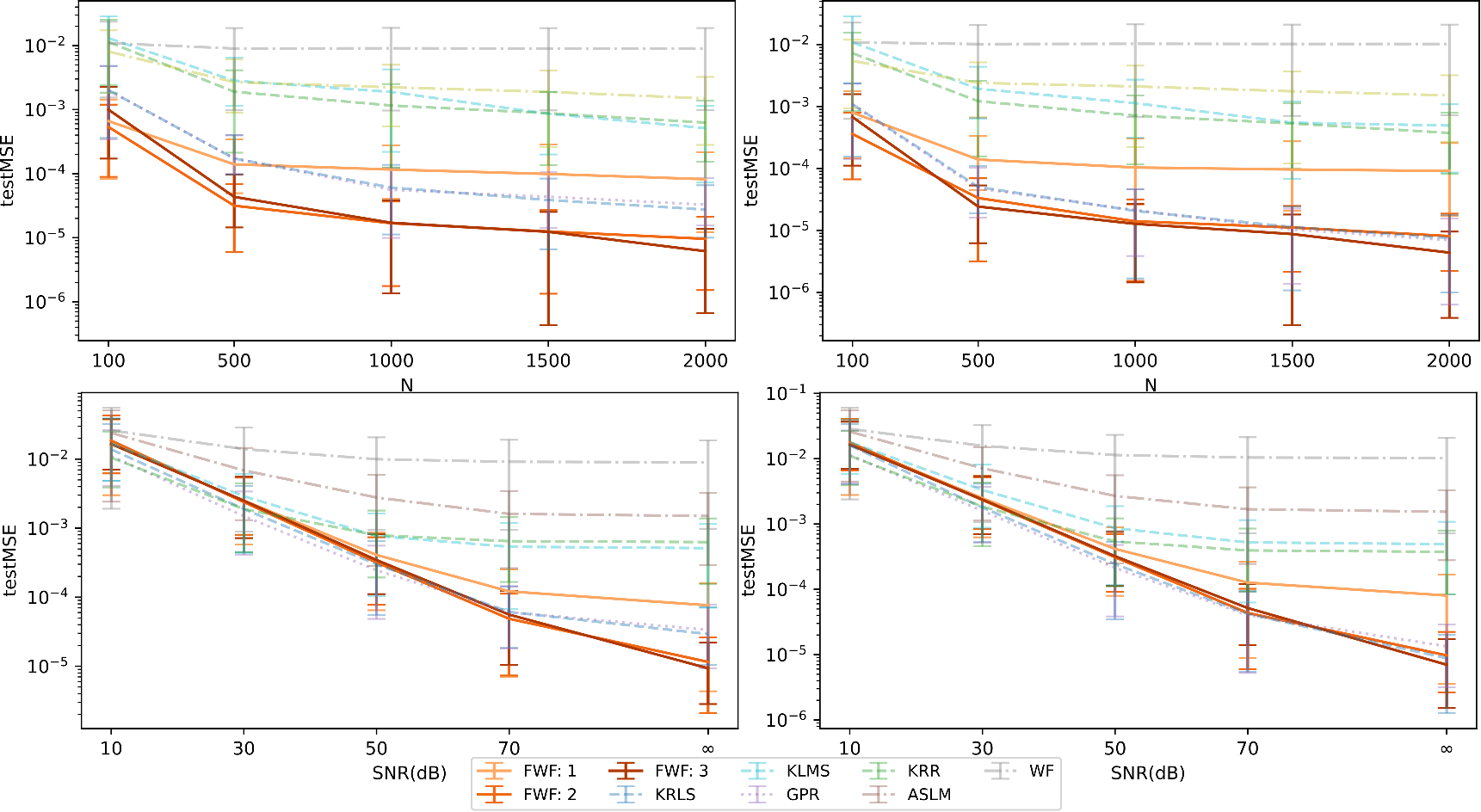}}
    \caption{(top)Comparison of Test MSE vs number of training samples(N) Mackey-Glass with window sizes of $L=20$(left) and $L=15$(right). (bottom)Comparison of Test MSE on noisy Mackey-Glass with window sizes of $L=20$(left) and $L=15$(right), $N=2000$ for all SNR levels. Error bars indicate best and worst case performance.}
    \label{fig:MG}
\end{figure*}

The top two plots in Figure \ref{fig:MG} show the test set MSE as a function of the number of training samples ($N$) for window sizes $L=15,20$. For each value of $N$ we give the average, best case, and worst case test MSE across five independent training and testing windows. We tested the FWF with sample embedding sizes of $D=1,2,3$. Increasing $D$ consistently improves the performance of the FWF. When compared to the other kernel methods, the FWF with $D=2,3$ is on par with KRLS and GPR for $L=15$, and outperforms KRLS and GPR for $L=20$. The FWF with $D=1$ plateaued after 500 samples, whereas larger sample embedding sizes required more data to converge. This supports our assertion of the trade-off between model capacity and data requirements for the estimation of the FWF solution. 

To assess robustness, we compare test MSE when additive white Gaussian noise at varying signal-to-noise ratios (SNR) corrupts the input to the model. The bottom two plots in \ref{fig:MG} show the performance for each method at each SNR.

\subsection{Lorenz Attractor}
The Lorenz attractor, introduced in \cite{Lorenz1963}, is a three-dimensional system of equations that can exhibit chaotic behavior.
\begin{equation}
    \frac{dx}{dt} = \sigma(y-x), \quad
    \frac{dy}{dt} = x(\rho-z)-y, \quad
    \frac{dz}{dt} = xy-\beta z 
\end{equation}
\begin{figure*}[h]
    \centering
    \makebox[\textwidth][c]{\includegraphics[width=1.\textwidth]{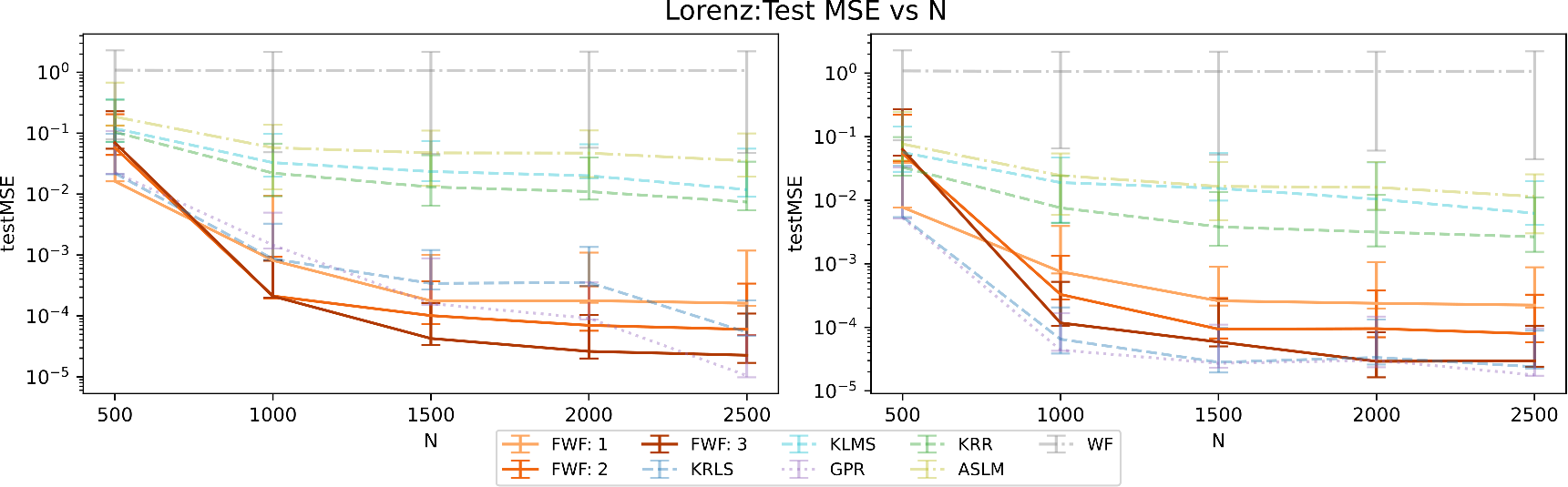}}
    \caption{Comparison of Test MSE on Lorenz task with window sizes of $L=20$(left) and $L=15$(right).}
    \label{fig:LO}
\end{figure*}

The Lorenz system is commonly used to test time series prediction and forecasting models (see \cite{Shahi2022},\cite{Liu2009}, and \cite{Chen2022}). For this experiment, the parameters $\sigma$, $\rho$, and $\beta$ are set to $10$, $28$, and $\frac{8}{3}$ respectively. The models are given the $x$-component of this system as input. The desired time series is the $z$-component of the Lorenz attractor five samples in the future. Each method is tested on five different training and testing windows for each value of N. As before, we perform a grid search ($0.5-2$ in increments of $0.25$) for the best kernel size for each method. Figure \ref{fig:LO} shows that the FWF with $D=3$ is as performant as KRLS and GPR for this task. The FWF again improves in performance with an increase in $D$.

\subsection{Sunspot Forecasting}
In this experiment, we test the methods on a real-world dataset (measured rather than simulated). The sunspot data \cite{SunSpot2017} contains monthly averages of the daily sunspot numbers reported from the WDC-SILSO, Royal Observatory of Belgium. The time series is standardized to have a mean of zero and a standard deviation of 1. The task for the models is to forecast the number of sunspots 10 samples in the future. The number of training samples is 2000, the number of test set samples is 300, and $L=10$ for all methods. The average MSE and standard deviation across 5 independent training and testing windows are given in Table \ref{tab:SunSpot}.  While the FWF with $D=3$ outperformed the other instantiations of the FWF in the previous experiments, on this task, it exhibited overfitting and large variance in test set performance, likely due to the increase in noise in this dataset. 
\begin{singlespace}
\begin{table}[h]
    \centering
    \small
    \begin{tabular}{|c|c|c|}
    \hline
    \textbf{Method} & \textbf{Train MSE} & \textbf{Test MSE} \\
    \hline
    FWF($D=1$) & $0.306 \pm 0.0059$   & $0.354 \pm 0.0080$ \\
    \hline
    FWF($D=2$) & $0.168 \pm 0.0016$ &  $\mathbf{0.182 \pm 0.0089}$ \\
    \hline
    FWF($D=3$) & $0.122 \pm 0.0014$   & $ 0.213 \pm 0.02$ \\
    \hline
    KRLS & $0.27 \pm 0.0012$   & $ 0.323 \pm 0.0072$ \\
    \hline
    KLMS & $0.366 \pm 0.0017$ & $0.374 \pm 0.0069$\\
    \hline 
    GPR& $0.279 \pm 0.0012$ & $0.324 \pm 0.0061$\\
    \hline
    KRR & $0.282 \pm 0.0011$ & $0.326 \pm 0.0054$ \\
    \hline
    ASLM & $0 \pm 0$ & $ 0.568 \pm 0.0093$ \\
    \hline
    WF & $0.33 \pm 0.0007 $ & $0.382 \pm 0.0047$\\
    \hline
    \end{tabular}
    \caption{Train and Test MSE for Sunspot forecasting on the normalized data.}
    \label{tab:SunSpot}
\end{table}
\end{singlespace}
\section{Conclusions}
In summary, we have successfully extended Parzen's MMSE in $\mathcal{H}_R$, to a nonlinear data-dependent RKHS, $\mathcal{H}_U$. By embedding the input random process statistics into the inner product of the space, the orthogonal projection of any desired r.v. (the MMSE solution) is immediately given by the cross-covariance function. Rather than implementing search techniques in a data-independent space, we build the RKHS in which our solution is defined, rendering explicit search unnecessary. Calculation of the FWF is simplified using an explicit finite-dimensional RKHS approximating the universal Gaussian kernel, where the $U$ operator becomes a finite-dimensional matrix defining the inner product in $\mathcal{H}_{\bm{U}}$. Experimentally, we demonstrate that when the assumptions made by the FWF are met, we outperform other kernel-based nonlinear adaptive filtering methods. Moreover, we can interpret the FWF solution as a nonlinear difference equation and extract the underlying functions learned by the FWF. 


Future work should address some practical limitations of the method. One of these limitations is that the dimensionality of $\mathcal{H}_{\bm{U}}$ grows exponentially with sample embedding size and truncation point. Other methods for explicit space approximations, perhaps ones that are more efficient in terms of dimensionality, should be explored. We should also expand on the interpretation of the FWF as a nonlinear difference equation, as this is unique to the FWF. Finally, one deeper question stemming from this work is: Can we define a practical technique for taking inner products in $\mathcal{H}_U$ without first referencing an extrinsic coordinate system? Achieving this would yield a fully data-determined Hilbert space which is truly the ``natural" setting for conditional expectations with respect to $X$.
\begin{singlespace}

\section*{Funding and AI Use}
\begin{small}
This material is based upon work supported by the Office of the Under Secretary of Defense for Research and Engineering under award number FA9550-21-1-0227, and partially supported by ONR grants N00014-21-1-2295 and N00014-21-1-2345. This research was supported by the SMART Scholarship Program. The authors used ChatGPT to identify spelling, grammar, and tense consistency issues. Each of these suggestions was reviewed by the authors, and the final edit was made.
\end{small}
\printbibliography

@book{haykin2014adaptive,
  title={Adaptive Filter Theory},
  author={Haykin, S.S.},
  isbn={9780132671453},
  lccn={2012025640},
  year={2014},
  publisher={Pearson}
}

@misc{cotter2011,
      title={{Explicit Approximations of the Gaussian Kernel}}, 
      author={Andrew Cotter and Joseph Keshet and Nathan Srebro},
      year={2011},
      eprint={1109.4603},
      archivePrefix={arXiv},
      primaryClass={cs.AI},
}

@article{Chen2022,
    author ={Chen, Y. and Qian, Y. and Cui, X.} ,
    title ={Time series reconstructing using calibrated reservoir computing} ,
    journal ={Sci Rep 12, 16318 (2022)} ,
    year = {2002}
}

@article{Duttweiler1973,
  author={Duttweiler, D. and Kailath, T.},
  journal={IEEE Transactions on Information Theory}, 
  title={An {RKHS} {A}pproach to {D}etection and {E}stimation {P}roblems--{IV}: {N}on-{G}aussian {D}etection}, 
  year={1973},
  volume={19},
  number={1},
  pages={19-28}}

@article{Engel2004,
  author={Engel, Y. and Mannor, S. and Meir, R.},
  journal={IEEE Transactions on Signal Processing}, 
  title={The {K}ernel {R}ecursive {L}east-{S}quares {A}lgorithm}, 
  year={2004},
  volume={52},
  number={8}}

@article{Hoerl2000,
 ISSN = {00401706},
 author = {Arthur E. Hoerl and Robert W. Kennard},
 journal = {Technometrics},
 number = {1},
 pages = {80--86},
 publisher = {[Taylor & Francis, Ltd., American Statistical Association, American Society for Quality]},
 title = {{Ridge Regression: Biased Estimation for Nonorthogonal Problems}},
 urldate = {2023-09-05},
 volume = {42},
 year = {2000}
}

@article{Kailath1975,
  title   = "An {RKHS} {A}pproach to {D}etection and {E}stimation {P}roblems-{P}art {II}:
             Gaussian {S}ignal {D}etection",
  author  = "Kailath, T and Weinert, H",
  journal = "IEEE Trans. Inf. Theory",
  volume  =  21,
  number  =  1,
  pages   = "15--23",
  year    =  1975
}

@ARTICLE{Kailath1971,
  author={Kailath, T.},
  journal={IEEE Transactions on Information Theory}, 
  title={An {RKHS} {A}pproach to {D}etection and {E}stimation {P}roblems--{I}: {D}eterministic {S}ignals in {G}aussian {N}oise},
  year={1971},
  volume={17},
  number={5},
  pages={530-549}}

@misc{Li2019,
      title={No-{T}rick ({T}reat) {K}ernel {A}daptive {F}iltering {U}sing {D}eterministic {F}eatures}, 
      author={Kan Li and Jose C. Principe},
      year={2019},
      eprint={1912.04530},
      archivePrefix={arXiv},
}

@book{Liu2010,
author = {Liu, Weifeng and Principe, Jose C. and Haykin, Simon},
title = {Kernel Adaptive Filtering: A Comprehensive Introduction},
year = {2010},
publisher = {Wiley Publishing},
edition = {1st}
}

@article{Liu2009,
  author={Liu, Weifeng},
  journal={IEEE Transactions on Signal Processing}, 
  title={{Extended Kernel Recursive Least Squares Algorithm}}, 
  year={2009},
  volume={57},
  number={10},
  pages={3801-3814},
  }

@article{Liu2008,
author = {Liu, Weifeng},
year = {2008},
month = {03},
pages = {543 - 554},
title = {The {K}ernel {L}east-{M}ean-{S}quare {A}lgorithm},
volume = {56},
journal = {IEEE Transactions on Signal Processing}
}

@article{Lorenz1963,
author = {Lorenz, Edward Norton},
year = {1963},
pages = {130-141},
title = {Deterministic nonperiodic flow},
volume = {20},
journal = {Journal of the Atmospheric Sciences},
}

@article{Mackey1977,
author = {Michael C. Mackey  and Leon Glass },
title = {{Oscillation and Chaos in Physiological Control Systems}},
journal = {Science},
volume = {197},
number = {4300},
pages = {287-289},
year = {1977},
}

@article{Mollenhauer2022,
  author  = {Mattes Mollenhauer and Stefan Klus and Christof Schütte and Péter Koltai},
  title   = {Kernel Autocovariance Operators of Stationary Processes: Estimation and Convergence},
  journal = {Journal of Machine Learning Research},
  year    = {2022},
  volume  = {23},
  number  = {327},
  pages   = {1--34},
}

@article{Muandet2017,
title={{Kernel Mean Embedding of Distributions: A Review and Beyond}},
volume={10},
ISSN={1935-8245},
number={1–2},
journal={Foundations and Trends in Machine Learning},
publisher={Now Publishers},
author={Muandet, Krikamol and Fukumizu, Kenji and Sriperumbudur, Bharath and Schölkopf, Bernhard},
year={2017},
pages={1–141} }

@article{Parzen1961,
author = {Emanuel Parzen},
title = {{An Approach to Time Series Analysis}},
volume = {32},
journal = {The Annals of Mathematical Statistics},
number = {4},
publisher = {Institute of Mathematical Statistics},
pages = {951 -- 989},
year = {1961},
}

@article{Parzen1959,
  author={Emanuel Parzen},
  title={{Statistical Inference on time series by Hilbert Space Methods}},
  journal={Technical Report},
  year={1959},
}

@article{Pokharel2006,
author = {Pokharel, P.P. and Xu, Jian-Wu and Erdogmus, Deniz and Principe, Jose},
year = {2006},
month = {06},
pages = {III - III},
title = {A {C}losed {F}orm {S}olution for a {N}onlinear {W}iener {F}ilter},
volume = {3},
journal = {Acoustics, Speech, and Signal Processing, 1988. ICASSP-88., 1988 International Conference on}
}

@book{Rasmussen2006,
  author = {Rasmussen, Carl Edward and Williams, Christopher K. I.},
  ee = {https://www.worldcat.org/oclc/61285753},
  publisher = {MIT Press},
  series = {Adaptive computation and machine learning},
  title = {Gaussian {P}rocesses for {M}achine {L}earning.},
  year = 2006
}

@article{Riesz1907,
  title   = "{Sur les Syst{\`e}mes Orthogonaux de Fonctions}",
  author  = "Riesz, Friedrich",
  journal = "Comptes rendus de l'Acad{\'e}mie des sciences",
  volume  =  144,
  pages   = "615--619",
  year    =  1907
}

@inproceedings{Rahimi2007,
 author = {Rahimi, Ali and Recht, Benjamin},
 booktitle = {Advances in Neural Information Processing Systems},
 publisher = {Curran Associates, Inc.},
 title = {{Random Features for Large-Scale Kernel Machines}},
 volume = {20},
 year = {2007}
}

@article{Stoica2004,
  author={Stoica, P. and Selen, Y.},
  journal={IEEE Signal Processing Magazine}, 
  title={{Model-Order Selection: A Review of Information Criterion Rules}}, 
  year={2004},
  volume={21},
  number={4},
  pages={36-47}}

@article{Santamaria2006,
  author={Santamaria, I. and Pokharel, P.P. and Principe, J.C.},
  journal={IEEE Transactions on Signal Processing}, 
  title={Generalized {C}orrelation {F}unction: {D}efinition, {P}roperties, and application to blind equalization}, 
  year={2006},
  volume={54},
  number={6},
  pages={2187-2197}}

@article{Shahi2022,
title = {Prediction of chaotic time series using recurrent neural networks and reservoir computing techniques: A comparative study},
journal = {Machine Learning with Applications},
volume = {8},
pages = {100300},
year = {2022},
issn = {2666-8270},
author = {Shahrokh Shahi and Flavio H. Fenton and Elizabeth M. Cherry}
}

@inbook{Shiryayev1992Interpolation,
	address = {Dordrecht},
	author = {Shiryayev, A. N.},
	booktitle = {Selected Works of A. N. Kolmogorov: Volume II Probability Theory and Mathematical Statistics},
	isbn = {978-94-011-2260-3},
	pages = {272--280},
	publisher = {Springer Netherlands},
	title = {Interpolation and Extrapolation of Stationary Random Sequences},
	year = {1992},
}

@inbook{Shiryayev1992sequences,
author={Shiryayev, A. N.},
chapter={Stationary Sequences in Hilbert Space},
title={Selected Works of A. N. Kolmogorov: Volume II Probability Theory and Mathematical Statistics},
year={1992},
publisher={Springer Netherlands},
pages={228-271},
}

@article{Hochreiter1997,
author = {Hochreiter, Sepp and Schmidhuber, J\"{u}rgen},
title = {Long {S}hort-{T}erm {M}emory},
year = {1997},
issue_date = {November 15, 1997},
publisher = {MIT Press},
address = {Cambridge, MA, USA},
volume = {9},
number = {8},
issn = {0899-7667},
doi = {10.1162/neco.1997.9.8.1735},
journal = {Neural Comput.},
month = nov,
pages = {1735–1780},
numpages = {46}
}

@misc{Lea2016,
      title={Temporal {C}onvolutional {N}etworks: {A} {U}nified {A}pproach to {A}ction {S}egmentation}, 
      author={Colin Lea and Rene Vidal and Austin Reiter and Gregory D. Hager},
      year={2016},
      eprint={1608.08242},
      archivePrefix={arXiv},
      primaryClass={cs.CV},
}

@misc{Miller2024,
      title={A {S}urvey of {D}eep {L}earning and {F}oundation {M}odels for {T}ime {S}eries {F}orecasting}, 
      author={John A. Miller and Mohammed Aldosari and Farah Saeed and Nasid Habib Barna and Subas Rana and I. Budak Arpinar and Ninghao Liu},
      year={2024},
      eprint={2401.13912},
      archivePrefix={arXiv},
      primaryClass={cs.LG},
}

@online{SunSpot2017,
    author = {Hathaway, David},
    title = {Sunspot Numbers},
    url  = {https://solarscience.msfc.nasa.gov/SunspotCycle.shtml},
    addendum = {accessed: 08/02/2024},
}

@book{Wahba1990,
author = {Wahba, Grace},
title = {Spline Models for Observational Data},
publisher = {Society for Industrial and Applied Mathematics},
year = {1990},
address = {},
edition   = {},
}

@INPROCEEDINGS{Vaerenbergh2013,
  author={Van Vaerenbergh, Steven and Santamaría, Ignacio},
  booktitle={2013 IEEE Digital Signal Processing and Signal Processing Education Meeting (DSP/SPE)}, 
  title={A {C}omparative {S}tudy of {K}ernel {A}daptive {F}iltering {A}lgorithms}, 
  year={2013},
  volume={},
  number={},
  pages={181-186}}

@book{Wiener1949,
    author = "Wiener, Norbert" ,
    title = "Extrapolation, Interpolation, and Smoothing of Stationary Time Series",
    publisher = "New York: Wiley" ,
    year = 1949
}

@inproceedings{Williams1995,
 author = {Williams, Christopher and Rasmussen, Carl},
 booktitle = {Advances in Neural Information Processing Systems},
 editor = {D. Touretzky and M.C. Mozer and M. Hasselmo},
 pages = {},
 publisher = {MIT Press},
 title = {{Gaussian Processes for Regression}},
 volume = {8},
 year = {1995}
}

@inproceedings{Yang2004nips,
 author = {Yang, Changjiang and Duraiswami, Ramani and Davis, Larry S},
 booktitle = {Advances in Neural Information Processing Systems},
 editor = {L. Saul and Y. Weiss and L. Bottou},
 pages = {},
 publisher = {MIT Press},
 title = {{Efficient Kernel Machines Using the Improved Fast Gauss Transform}},
 volume = {17},
 year = {2004}
}

@article{Qine2020,
  author={Qin, Zhengda},
  journal={IEEE Transactions on Signal Processing}, 
  title={Augmented Space Linear Models}, 
  year={2020},
  volume={68},
  number={},
  pages={2724-2738}}

@article{Wang2022,
title = {A {K}ernel {R}ecursive {M}inimum {E}rror {E}ntropy {A}daptive {F}ilter},
journal = {Signal Processing},
volume = {193},
pages = {108410},
year = {2022},
issn = {0165-1684},
doi = {https://doi.org/10.1016/j.sigpro.2021.108410},
author = {Gang Wang and Xinyue Yang and Lei Wu and Zhenting Fu and Xiangjie Ma and Yuanhang He and Bei Peng}
}

@article{He2023,
title = {Generalized {M}inimum {E}rror {E}ntropy for {R}obust {L}earning},
journal = {Pattern Recognition},
volume = {135},
pages = {109188},
year = {2023},
issn = {0031-3203},
author = {Jiacheng He and Gang Wang and Kui Cao and He Diao and Guotai Wang and Bei Peng}
}

@article{Wang2025,
title = {An {E}fficient {K}ernel {A}daptive {F}iltering {A}lgorithm with {A}daptive {A}lternating {F}iltering {M}echanism},
journal = {Digital Signal Processing},
volume = {159},
pages = {104997},
year = {2025},
issn = {1051-2004},
doi = {https://doi.org/10.1016/j.dsp.2025.104997},
author = {Hong Wang and Hongyu Han and Sheng Zhang and Jinhua Ku}
}

@article{Vega2019,
title = {{Learning from Data Streams Using Kernel Least-Mean-Square with Multiple Kernel-Sizes and Adaptive Step-size}},
journal = {Neurocomputing},
volume = {339},
pages = {105-115},
year = {2019},
issn = {0925-2312},
author = {Sergio Garcia-Vega and Xiao-Jun Zeng and John Keane}
}

@article{Sherstinsky2020,
title = {Fundamentals of {R}ecurrent {N}eural {N}etwork ({RNN}) and {L}ong {S}hort-{T}erm {M}emory ({LSTM}) {N}etwork},
journal = {Physica D: Nonlinear Phenomena},
volume = {404},
pages = {132306},
year = {2020},
issn = {0167-2789},
author = {Alex Sherstinsky}
}

@misc{Wen2023,
      title={Transformers in {T}ime {S}eries: {A} {S}urvey}, 
      author={Qingsong Wen and others},
      year={2023},
      eprint={2202.07125},
      archivePrefix={arXiv},
      primaryClass={cs.LG},
}
\end{singlespace}
\appendix
\section{The $\mathbf{U}$ matrix}\label{A1}
In section \ref{sec:MMSEComputation} an abbreviated construction of the $\mathbf{U}$ matrix is given. Here we show a more explicit construction of this matrix as a block matrix. Due to strict stationarity, we can remove the dependence on $s,t$ and let $s = t - \tau $. Then $\bm{U}(\tau, n, m) = \mathbb{E}[\phi_n(\bm{X}_t^D)\phi_m(\bm{X}_{t-\tau}^D)]$.
Intuitively, $\bm{U}(\tau, n, m)$ measures the correlation between the projection of $\bm{X}_t^D$ and $\bm{X}_{t-\tau}^D$ across all dimensions of $\mathcal{H}_S$. For each value of $\tau=0, 1, \dots, L-1$, we can store $\bm{U}(\tau,n,m)$ in a $M \times M$ dimensional matrix, $\bm{U}_\tau = \mathbb{E}[\phi(\bm{X}_t^{D})\phi(\bm{X}^{D}_{t-\tau})^\top]$.
Finally, the matrix $\mathbf{U}$ is a $((M\cdot L)\times(M\cdot L))$ dimensional positive semi-definite matrix, 
\begin{equation}
     \mathbf{U}= \begin{bmatrix}
         \mathbf{U}_0 & \mathbf{U}_{1}& ... & \mathbf{U}_{L-1}\\
         \mathbf{U}_{1}^{\top} & \mathbf{U}_0 &\cdots & \mathbf{U}_{L-2}^{\top} \\
         \vdots & \vdots &\ddots & \vdots\\ 
         \mathbf{U}_{L-1}^{\top}&\mathbf{U}_{L-2}^{\top} &\cdots& \mathbf{U}_0
     \end{bmatrix}
\label{Ucalculation}
\end{equation}
Since $\mathbf{U}$ is the covariance matrix of $\phi(\bm{X}_t^{DL})$ (the same for all $t$ because of stationarity), it is a positive semi-definite matrix \cite{Muandet2017}. The $\mathbf{U}$ matrix describes the correlations of the projection of $\bm{X}^{DL}$ across both space (the features of $\mathcal{H}_S$) and time. Assuming strict stationarity, the following relationship between the elements of the submatrices of $\mathbf{U}$ and $U(\tau,x,y)$ is given as, 
\begin{equation}
    U(\tau,x,y) \approx \sum_{n=1}^{M} \sum_{m=1}^{M} \bm{U}(\tau, n,m)\phi_n(x)\phi_m(y).
\label{Umatrix2op}
\end{equation}


\end{document}